# Model-Free Reinforcement Learning for Financial Portfolios: A Brief Survey

Yoshiharu Sato [†]

## Abstract

Financial portfolio management is one of the problems that are most frequently encountered in the investment industry. Nevertheless, it is not widely recognized that both Kelly Criterion and Risk Parity collapse into Mean Variance under some conditions, which implies that a universal solution to the portfolio optimization problem could potentially exist. In fact, the process of sequential computation of optimal component weights that maximize the portfolio's expected return subject to a certain risk budget can be reformulated as a discrete-time Markov Decision Process (MDP) and hence as a stochastic optimal control, where the system being controlled is a portfolio consisting of multiple investment components, and the control is its component weights. Consequently, the problem could be solved using model-free Reinforcement Learning (RL) without knowing specific component dynamics. By examining existing methods of both value-based and policy-based model-free RL for the portfolio optimization problem, we identify some of the key unresolved questions and difficulties facing today's portfolio managers of applying model-free RL to their investment portfolios.

[†] E-mail: yoshi2233@protonmail.ch. Web: yoshi2233.strikingly.com

# 1. Introduction

Reinforcement Learning (RL) [1] produces autonomous agents that interact with their environments with optimal behaviors that are learned through trial and error. Rapid advancements in Deep Neural Networks (DNNs) in the past several years has allowed RL to solve decision-making problems with high-dimensional state-action spaces, thereby establishing the field of Deep Reinforcement Learning (DRL), which has been widely successful in playing video games and board games. The Deep Q-Network (DQN) [2] and its various extensions (e.g., *Rainbow* [3]) could learn to play a range of Atari 2600 games at a superhuman level by merely observing screen pixels. In recent years, a hybrid DRL algorithm called *AlphaGo* [4] defeated a human world champion in the game of Go, and its more advanced and generalized version, *AlphaZero* [5], convincingly defeated world champion programs in chess, Go, and Shogi, given no domain knowledge other than the underlying rules.

Largely inspired by these successful DRL algorithms, there in recent years have been an increasing number of published literature on applying RL to dynamic financial decision-making problems. For instance, Gervais et al. [6] construct a Markov Decision Process (MDP) and optimize adversarial strategies (double spending and selfish mining [7]) against Proof-of-Work (PoW) blockchains [8] through policy iteration. Halperin [9] constructs an MDP for a discrete-time version of the Black-Scholes-Merton (BSM) model [10][11] and demonstrates the optimal hedging and pricing of stock options using model-free Q-Learning [12]. Buehler et al. [13] presents a DRL framework to hedge a portfolio of derivatives under transaction costs, where the framework does not depend on specific market dynamics. Jiang et al. [14] use the model-free Deep Deterministic Policy Gradient (DDPG) [15] to dynamically optimize cryptocurrency portfolios. Similarly, Liang et al. [16] optimize stock portfolios by using the DDPG as well as the Proximal Policy Optimization (PPO) [17].

Dynamic portfolio optimization is indeed one of the problems which investment industry practitioners most frequently encounter. Its three major paradigms are 1) Mean Variance, 2) Kelly Criterion, and 3) Risk Parity. The Mean Variance computes the Efficient Frontier (EF) defined as the set of investments that yield the highest achievable mean excess return for any given level of risk. The Kelly Criterion maximizes the expected geometric growth rate of a portfolio. The Risk Parity equalizes risk from different portfolio components by having weights that are inversely proportional to the component's return volatility. In actuality, Kelly Criterion is a special case of Mean Variance, and Risk Parity collapses into Mean Variance under some conditions on return correlation and Sharpe ratio. These two facts imply that the problem of portfolio optimization could have some universal solution. The problem is a dynamic and intertemporal process of determining the optimal portfolio weights (fractions of capital allocation on component investments) that maximize the portfolio's expected return subject to a certain risk budget. Uncertainty about future market states (i.e., investment returns can hardly be predicted sequentially with a sufficient accuracy) makes it a stochastic optimal control problem in continuous state and action spaces – a problem which could be solved by model-free RL.

This paper presents a brief survey on both value-based and policy-based model-free RL methods for the portfolio optimization problem. By examining existing methods for the problem, we identify some of the key unresolved questions and difficulties facing today's portfolio managers of applying model-free RL to their investment portfolios. The paper is organized as follows. In Chapter 2 we review the three major portfolio paradigms and derive their equivalence. In Chapter 3 we take a look at various model-free RL methods applied to the portfolio optimization problem. In Chapter 4 we conclude by making a detailed discussion on the issues of model-free RL.

# 2. Portfolio Optimization

In this chapter, we first provide a brief overview of the three major paradigms in portfolio optimization – i.e., Mean Variance, Kelly Criterion, and Risk Parity. We subsequently demonstrate that the latter two both collapse to the Mean Variance by deriving the Markowitz-Kelly equivalence and the Markowitz-RP equivalence.

## 2.1. Mean Variance

Markowitz' Modern Portfolio Theory (MPT) [19] is the dominant paradigm in portfolio optimization. It consists in computing the Mean-Variance Efficient Frontier (EF) defined as the set of investments that yield the highest achievable mean excess return with respect to the risk-free rate for any given level of risk measured in terms of standard deviation. Specifically, consider a universe of $n$ investments with returns $x$ and their mean $\mu$, standard deviaion $\sigma$, covariance matrix $\Sigma = [\sigma_{ij}]$ (where $\sigma_{ii} = \sigma_i^2$ and $\sigma_{ij} = \rho_{ij}\sigma_i\sigma_j$ for $i \neq j$), and the portfolio weight vector $\omega$. The solution to the *unconstrained* portfolio optimization problem (i.e., full investment $\Sigma\omega_i = 1$ with target mean $\mu_p$) is obtained by minimizing the Lagrange function w.r.t. $\omega$:

$$L(\omega, \gamma, \lambda) = \frac{1}{2}\omega^T \Sigma \omega - \gamma(\omega^T \mathbf{1}_n - 1) - \lambda(\omega^T \mu - \mu_p) \qquad (1)$$

where $\gamma$ and $\lambda$ are multipliers, $\omega^T\Sigma\omega$ is the portfolio variance $Var(\omega^T x)$, and $\mathbf{1}_n$ is a vector of $n$ ones. See [22] for the analytical solution, which is unfortunately not practical because the obtained weights are highly unstable and can be negative values.

For the inequality-constrained optimization problem (to which no analytical solution exists), Markowitz developed the Critical Line Algorithm (CLA) [20][21], which not only optimizes general quadratic functions subject to linear inequality constraints but also guarantees that the exact numerical solution is found after a finite number of iteration along with the entire EF. Using the CLA it is possible to construct the Minimum Variance Portfolio (MVP), which is the leftmost portfolio of the constrained EF (not allowing short selling), as well as the Maximum Sharpe Ratio Portfolio (MSRP; a.k.a. Tangency Portfolio)[1].

However, CLA solutions are known to be unstable since small deviations in return forecasts will cause the algorithm to result in wildly different portfolios [23]. This is because the precision matrix, or the inverted covariance matrix, is prone to large errors when the covariance matrix is numerically ill-conditioned (i.e., having a high condition number) [24]. López de Prado [25] provides a detailed discussion on this issue. Namely, the condition number of the covariance matrix grows as we add correlated, multicollinear investments into the portfolio, which number becomes so high at some point that numerical errors make the precision matrix too unstable to the point where a small change on any entry in the covariance matrix will lead to a very different inverse. Consequently, the stronger is the multicollinearity among the portfolio components, the higher is the condition number of the covariance matrix, hence the more unstable is the precision matrix. The CLA algorithm is therefore likely to fail precisely when there is a greater need for finding a diversified portfolio (a.k.a. Markowitz' Curse). Various solutions to this issue have been proposed (see [25]) but we would not go into any detail in this paper.

---

1 A Python implementation of the CLA algorithm for MVP and MSRP portfolio construction is made publicly available by Bailey & López de Prado [26].

## 2.2. Kelly Criterion

Another major paradigm in portfolio optimization is the Kelly Criterion [27][28], which consists in maximizing the expected logarithm of the terminal wealth (or the median thereof [29]) of an investment strategy, providing the optimal per-trade position size that maximizes long-term geometric growth for repeated trades over time. It also minimizes the expected time to reach a given wealth target, and asymptotically dominates all essentially different strategies [30][31]. It takes the simple algebraic form:

$$k = \left(\frac{p}{a}\right) - \left(\frac{1-p}{b}\right) \qquad (2)$$

where $k \in [0,1]$ is the optimal position size of the strategy, $p$ is the probability of winning per trade, $a$ is the expected net loss per trade, and $b$ is the expected net gain per trade (see [32] for the derivation). An import implication here is that the value of $k$ does not depend on the total number of trades made by the strategy, thereby making it myopic[2].

A multivariate and model-free Kelly Criterion Portfolio (KCP) [35] considers $n$ investments in the portfolio. For each investment indexed by $i = 1, ..., n$ we denote its return by $x_i$ and wealth fraction by $\omega_i$ (we start with total wealth of 1). The residual wealth $(1 - \Sigma\omega_i)$ is invested in a risk-free asset with return $r_f$. The KCP portfolio is then defined as the weight vector $\boldsymbol{\omega}_{KCP}$ such that:

$$\boldsymbol{\omega}_{KCP} = \underset{\omega \in \mathbb{R}^n}{\operatorname{argmax}} \operatorname{E}\left[\log\left((1+r_f) + \sum_{i=1}^{n} \omega_i(x_i - r_f)\right)\right] \qquad (3)$$

which can be solved as a quadratic optimization problem after expanding it as the Taylor series around $\boldsymbol{\omega_0} = [0, ..., 0]^T$ (see [35] for the unconstrained solution). The constrained solution (i.e., no leverage and no short selling) can be obtained using a numerical optimizer.

KCP portfolios often result in a very risky short-term behavior because of logarithm's lack of risk aversion, and consequently large concentrated investments in the portfolio, hence they can have considerable losses at least in the short term [36]. Due to said non-diversification and overconcentration, the KCP is therefore not commonly used by portfolio managers in the investment industry except for those looking for superior long-term returns. Circumstantial evidence suggests that George Soros and Warren Buffett are Kelly investors since their portfolios are concentrated in few investments. In order to protect KCP portfolios from bad-scenario outcomes, it is advised to lower the amount of bets and diversify across multiple uncorrelated investments.

## 2.3. Markowitz-Kelly Equivalence

Laureti et al. [37] show that the KCP lies on the constrained EF when mean returns and volatilities of the portfolio components are small and there is no risk-free asset (see Figure 1). This suggests that the KCP is a special case of Markowitz portfolios. In fact, Markowitz himself states that "on the EF there is a point which approximately maximizes E[ln($W_1$)]" [38].

---

2 In fact, the value of $k$ remains constant if the true values of $p$, $a$ and $b$ are known a priori [33][34].

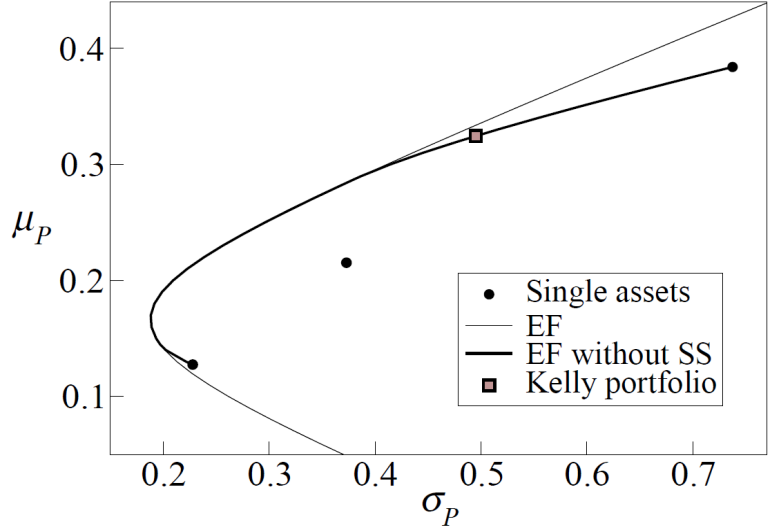

**Figure 1** (extracted from p.7 of [37]): Efficient Frontier, constrained Efficient Frontier (no short selling), and KCP portfolio for three risky assets.

To illustrate the similarity between the two portfolios, we now arrange Markowitz' portfolio optimization problem in Section 2.1 using a risk-aversion constant $\lambda$, which is the amount of additional expected return an investor requires to accept additional risk (i.e., the second central moment):

$$\mathrm{E}\left[\boldsymbol{\omega}^T \boldsymbol{x}\right] - \lambda \operatorname{Var}\left[\boldsymbol{\omega}^T \boldsymbol{x}\right] = \boldsymbol{\mu}^T \boldsymbol{\omega} - \frac{\delta}{2} \boldsymbol{\omega}^T \boldsymbol{\Sigma} \boldsymbol{\omega} \qquad (4)$$

where $\delta = 2\lambda$ [39]. Since $\operatorname{Var}(\boldsymbol{\omega}^T\boldsymbol{x}) = \mathrm{E}[(\boldsymbol{\omega}^T\boldsymbol{x})^2] - \mathrm{E}[\boldsymbol{\omega}^T\boldsymbol{x}]^2$ by definition and $\mathrm{E}[\boldsymbol{\omega}^T\boldsymbol{x}] < 1$ in most practical cases, we can assume that $\mathrm{E}[\boldsymbol{\omega}^T\boldsymbol{x}] \gg \mathrm{E}[\boldsymbol{\omega}^T\boldsymbol{x}]^2 \sim 0$, and hence rewrite the l.h.s. of Eq. 4 as:

$$\mathrm{E}\left[\boldsymbol{\omega}^T \boldsymbol{x}\right] - \lambda \mathrm{E}\left[\left(\boldsymbol{\omega}^T \boldsymbol{x}\right)^2\right]. \qquad (5)$$

Now, we consider the KCP with a zero risk-free rate and rewrite Eq. 3 as:

$$\boldsymbol{\omega}_{KCP} = \underset{\boldsymbol{\omega} \in \mathbb{R}^n}{\operatorname{argmax}} \; \mathrm{E}\left[\log\left(1 + \boldsymbol{\omega}^T \boldsymbol{x}\right)\right]. \qquad (6)$$

Using the following Taylor series formula [40] for $|x| \leqq 1$ and $x \neq -1$:

$$\log(1+x) = \sum_{k=1}^{\infty} (-1)^{k+1}\left(\frac{x^k}{k}\right) \qquad (7)$$

we can rewrite $\mathrm{E}[\log(1 + \boldsymbol{\omega}^T\boldsymbol{x})]$ in Eq. 6 as:

$$\mathrm{E}\left[\sum_{k=1}^{\infty}(-1)^{k+1}\frac{\left(\boldsymbol{\omega}^T\boldsymbol{x}\right)^k}{k}\right] \approx \mathrm{E}\left[\sum_{k=1}^{2}(-1)^{k+1}\frac{\left(\boldsymbol{\omega}^T\boldsymbol{x}\right)^k}{k}\right]. \qquad (8)$$

Expanding the r.h.s. of Eq. 8 we get,

$$\mathrm{E}\left[(-1)^{1+1}\frac{(\boldsymbol{\omega}^T\boldsymbol{x})^1}{1}+(-1)^{2+1}\frac{(\boldsymbol{\omega}^T\boldsymbol{x})^2}{2}\right]$$
$$=\mathrm{E}\left[\frac{(\boldsymbol{\omega}^T\boldsymbol{x})^1}{1}+(-1)\frac{(\boldsymbol{\omega}^T\boldsymbol{x})^2}{2}\right] \quad (9)$$
$$=\mathrm{E}[\boldsymbol{\omega}^T\boldsymbol{x}]-\frac{1}{2}\mathrm{E}[(\boldsymbol{\omega}^T\boldsymbol{x})^2]$$

which is equivalent to Markowitz portfolio (Eq. 5) with $\lambda = 0.5$ [41]. We would call this the Markowitz-Kelly equivalence of optimum portfolio for weakly risk-averse investors, and define the Markowitz-Kelly Portfolio (MKP) $\boldsymbol{\omega}_{MKP}$ as:

$$\boldsymbol{\omega}_{MKP} = \underset{\boldsymbol{\omega} \in \mathbb{R}^n}{\mathrm{argmax}}\, \mathrm{E}[\boldsymbol{\omega}^T\boldsymbol{x}]-\frac{1}{2}\mathrm{E}[(\boldsymbol{\omega}^T\boldsymbol{x})^2]. \quad (10)$$

To reiterate, the MKE has three assumptions:
1. The investor is weakly risk-averse ($\lambda = 0.5$).
2. Expected arithmetic daily return is less than 100% (i.e., $\mathrm{E}[\boldsymbol{\omega}^T\boldsymbol{x}] < 1$).
3. Second-order Taylor series sufficiently approximates $\log(1 + \boldsymbol{\omega}^T\boldsymbol{x})$.

We suppose the first assumption is given and the second one will empirically hold in many investments. The third one will be violated if the portfolio generates high alpha (i.e., $\mathrm{E}[\boldsymbol{\omega}^T\boldsymbol{x}] > 0$) and its daily returns are positively skewed, in other words their probability density function has a fat right tail (i.e., $\mathrm{E}[(\boldsymbol{\omega}^T\boldsymbol{x})^3] > 0$) [41]. Nevertheless, as David E. Shaw said "quantitative trading became more challenging with every passing year" [42], we suppose that the third assumption also holds in many cases.

  We believe that the MKP portfolios are not particularly practical because of the fact that $\mathrm{E}[\boldsymbol{\omega}^T\boldsymbol{x}] \gg 0.5\mathrm{E}[(\boldsymbol{\omega}^T\boldsymbol{x})^2]$, in other words they effectively maximize the arithmetic mean. Arithmetic mean maximization is optimal when the investor is restricted to bet only once or bet the same absolute amount repeatedly without reinvesting profits [27], which is not the case for the portfolio optimization problem.

## 2.4. Risk Parity

Risk Parity (RP) is another major paradigm in portfolio management that gained popularity mainly among high risk-aversion investors after the 2008 financial crisis, during which the S&P500 index lost approximately 50% of its value. The idea is simply to equalize risks or return volatilities of portfolio components. One of the advantages is that it does not have a sensitivity issue like the return estimation errors in Markowitz portfolios as we discussed in Section 2.1, since RP only requires estimation of component return volatility, which is more robust than return forecasts [43].

  There are mainly two approaches in this paradigm: the Equal Stand-alone Risk (ESR) and the Equal Risk Contribution (ERC). The ESR portfolios equalize the products of the component weights and the component return volatilities, making the weights inversely proportional to the volatilities:

$$\omega_i = \frac{\sigma_i^{-1}}{\sum_{j=1}^{N} \sigma_j^{-1}} \quad (11)$$

whereas the ERC portfolios equalize the contribution of each component to the portfolio volatility:

$$RC_i = \omega_i \frac{(\Sigma \boldsymbol{\omega})_i}{\sqrt{\boldsymbol{\omega}^T \Sigma \boldsymbol{\omega}}} \tag{12}$$

to which no analytical solution exists. The ERC can be interpreted alternatively as a portfolio in which component weights are inversely proportional to their betas in regards to the portfolio itself [44]. Both the ESR and ERC require the use of leverage in order to achieve a desired return or risk target, which potentially increases downside risk and also can increase transaction costs due to exacerbated turnover.

## 2.5. Markowitz-RP Equivalence

Markowitz' portfolio optimization problem in Section 2.1 can also be rearranged using the Constant Relative Risk Aversion (CRRA) utility:

$$U(W) = \frac{W^{1-\gamma} - 1}{1 - \gamma} \tag{13}$$

where $W$ is the wealth of the investor and $\gamma > 0$ is the constant risk aversion ($\gamma = 0$ means risk neutral). Let us assume risky asset's price follows a standard Geometric Brownian Motion (GBM) with parameters ($\mu$, $\sigma$), and the portfolio is composed of one risky asset (with Sharpe ratio $s$) and one risk-free asset. The optimal allocation to the risky asset is then defined as: $\omega^* = (\mu - r_f) / \gamma\sigma^2 = s / \gamma\sigma$ (see [18]).

Now, let $\boldsymbol{\sigma} = [\sigma_1, \ldots, \sigma_N]^T$ be the vector of standard deviation of $N$ risky assets, $D = diag(\boldsymbol{\sigma})$ be the $N \times N$ diagonal matrix with $\boldsymbol{\sigma}$ as its diagonal elements, and $C$ be the correlation matrix of the assets, so $\Sigma = DCD$. Let us assume further that all the risky assets have the same constant Sharpe ratio $k = (\mu - r_f) / \sigma$, and that all the pairwise correlations are the same (i.e., $\rho_{ij} = \rho$ for $i \neq j$), which implies: $C\mathbf{1}_N = [1 + (N-1)\rho]\mathbf{1}_N$. Therefore,

$$C^{-1}\mathbf{1}_N = \frac{1}{1+(N-1)\rho}\mathbf{1}_N \tag{14}$$

and we get:

$$\begin{aligned} D\boldsymbol{\omega}^* &= D\left[\frac{1}{\gamma}\Sigma^{-1}(\mu - r_f)\right] = \frac{1}{\gamma}D(D^{-1}C^{-1}D^{-1})k\sigma \\ &= \frac{k}{\gamma}C^{-1}\mathbf{1}_N = \frac{k}{\gamma[1+(N-1)\rho]}\mathbf{1}_N. \end{aligned} \tag{15}$$

This means:

$$\omega_i^* = \frac{k}{\gamma[1+(N-1)\rho]\sigma_i}, \forall i \tag{16}$$

which is equivalent to the Markowitz portfolio with equal stand-alone risk and equal risk contribution [18]. We would call this the Markowtiz-RP equivalence under identical Sharpe ratios and pairwise correlations (obviously the assumptions are purely hypothetical).

# 3. Model-Free Reinforcement Learning

The Markowitz-Kelly equivalence and the Markowtiz-RP equivalence give us a great implication, which is that the problem of portfolio optimization could have some universal solution. It is a dynamic and intertemporal process of determining the optimal portfolio weights that maximize the portfolio's expected return subject to a certain risk budget. Uncertainty about future markets states makes it a stochastic optimal control problem in continuous state and action spaces – a problem which could be solved by model-free Reinforcement Learning (RL). In this chapter, we discuss some of the existing methods of both value-based and policy-based model-free RL for the portfolio optimization problem.

## 3.1. Overview

One can postulate that the portfolio optimization problem can be reformulated as a discrete-time (partially observable) Markov Decision Process (MDP) and hence as a stochastic optimal control, where the system being controlled in discrete time is a portfolio consisting of multiple investments, and the control is the portfolio weights (fractions of capital allocation). The problem is then solved by a sequential maximization of portfolio returns as rewards in a Bellman optimality equation[3]. If the MDP is fully deterministic (or state transition probabilities are known) and if a reward function is also known, the Bellman optimality equation can be solved using a recursive backward value iteration method of Dynamic Programming (DP). If, on the other hand, the system dynamics is unknown and the optimal policy should be computed from samples, one can use model-free Reinforcement Learning (RL) to solve the problem. In portfolio optimization, neither the future returns of investments nor the state transition probabilities are known. Consequently, the MDP is nondeterministic and one can use RL for the problem. With model-free RL methods, no model of the investment returns is required because a Bellman optimality equation can be solved approximately without any knowledge of the underlying dynamics but relying solely on sample data.

Let us consider the standard RL setting for the portfolio optimization problem. At each time step $t$, an agent observes a current state $s_t \in S$ and chooses an action $a_t \in A$ according to its policy $\pi$. The agent subsequently observes the next state $s_{t+1}$ and receives a scalar reward $r_t = r(s_t, a_t)$.

$$R_t = \sum_{k=t}^{\infty} \gamma^{k-t} r(s_k, a_k) \tag{17}$$

is the total cumulative return from time step $t$ onwards with a discount factor $\gamma \in (0,1]$. For the portfolio optimization problem, $R_t$ is often replaced with undiscounted cumulative wealth $W_t$ with some initial wealth $W_0$:

$$W_t = W_0 \prod_{k=1}^{t} (1+r_k). \tag{18}$$

The state value $V^\pi(s) = E[R_t|s_t=s; \pi]$ is the expected return for following policy $\pi$ in state $s$. In a similar manner, the state-action value or Q-value $Q^\pi(s, a) = E[R_t|s_t=s, a; \pi]$ is the expected return for selecting action $a$ in state $s$ and following policy $\pi$. The optimal state-action value is then defined as $Q^*(s, a) = \max_\pi Q^\pi(s, a)$, and the optimal action as $a^* = \text{argmax}_a Q^\pi(s, a)$. The goal of the agent is to maximize the expected return from the start state, denoted by the performance objective: $J(\pi) = E[R_0|\pi]$. The optimal policy,

---

3 If rewards are not penalized by volatility, it would resemble a risk-neutral investor ($\lambda = 0$).

which is the goal to obtain in RL, is therefore defined as: $\pi^* = \text{argmax}_\pi J(\pi)$.

RL makes use of the recursive relationship known as the Bellman equation:

$$Q^\pi(s_t, a_t) = \mathrm{E}\left[r(s_t, a_t) + \gamma \mathrm{E}\left[Q^\pi(s_{t+1}, a_{t+1})\right]\right] \tag{19}$$

which can be rewritten without the inner expectation if the target policy is deterministic ($\mu: S \rightarrow A$):

$$Q^\mu(s_t, a_t) = \mathrm{E}\left[r(s_t, a_t) + \gamma Q^\mu(s_{t+1}, a_{t+1})\right]. \tag{20}$$

$Q^\mu$ can be learned *off-policy*, using transitions generated from some different stochastic behavior policy.

## 3.2. Value-Based Methods

In model-free and value-based RL methods (such as Q-Learning [12]), the agent's action is determined not by finding the optimal policy $\pi^*$ directly but instead obtaining the optimal state-action value function $Q^*(s, a)$, whose convergence is guaranteed in Q-Learning within both deterministic and stochastic frameworks [45][46]. $Q^*(s, a)$ is often approximated with a neural network because of its capability of universal approximation and representation learning (which is ideal for large state and action spaces from which Bellman's curse of dimensionality arises). The use of a neural network is also more efficient than a single Q-table because the network can generalize from states already encountered by the agent and therefore reduce the memory usage and the amount of computation. However, introducing nonlinear function approximators means that convergence is no longer guaranteed.

Q-Learning aims to directly approximate the optimal state-action value function: $Q^*(s, a) \sim Q(s, a; \theta)$, where $Q(s, a; \theta)$ is an approximate state-action value function with parameter vector $\theta$. In one-step Q-Learning, the parameters are learned by iteratively minimizing a sequence of loss functions, the $i$-th of which function is defined as:

$$L_i(\theta_i) = \mathrm{E}\left[r + \gamma \max_{a'} Q(s', a'; \theta_{i-1}) - Q(s, a; \theta_i)\right]^2 \tag{21}$$

where $s'$ is the next state immediately observed by the agent after state $s$. Q-Learning is off-policy because $Q$ is not updated by transitions generated by the behavioral policy (i.e., the policy directly derived from $Q$ itself).

Deep Q-Network (DQN) [2] uses a Deep Neural Network (DNN), originally a Convolutional Neural Network (CNN), for the learning of low-dimensional feature representations of high-dimensional states and also for the function approximation of the state-action value function $Q^\pi(s, a)$ for all discrete actions. The DQN agent learns based on an $\varepsilon$-greedy exploration strategy where it chooses a random action with probability $1 - \varepsilon$, using random mini-batches sampled from an experience replay buffer, which is a finite-sized cache that stores sample transitions experienced by the agent: ($s_t$, $a_t$, $r_t$, $s_{t+1}$), in order to remove sample correlations. After the learning, the agent can choose the optimal action $a^*$ in a single forward pass of the network, but it cannot be straightforwardly applied to continuous domains since finding $a^*$ in a continuous space requires an iterative optimization process at every time step. Naively discretizing the space would suffer from the curse of dimensionality – i.e., the number of discrete actions increases exponentially with the number of degrees of freedom.

Du et al. [47] use Q-Learning (without a neural network) in discretized market states to optimize a portfolio of a riskless asset (cash) and a risky asset (stock market portfolio) with transaction costs considered at each rebalancing period. Three state-action value functions are constructed using different types of investment performance metrics; namely cumulative profit, Sharpe ratio, and Differential Sharpe Ratio (DSR) [48]. DSR can be considered as the marginal utility in terms of how much risk the investor has a willingness to bear for one unit increment of the Sharp ratio. The authors find that the portfolio's performance varies significantly depending on the type of the state-action value function, and show that the portfolio using the DSR function achieves the best performance. However, they do not specifically mention if the result was obtained in out-of-sample testing. The largely varied portfolio performance may be attributable to the fact that 1) determining the global optimal policy in an arbitrary Q-value function is often infeasible without guaranteed convexity, and 2) Q-Learning suffers from instability with noisy datasets for optimal policy selection. The authors conclude that policy-based RL methods, which we will describe in the next section, is more stable and efficient than value-based RL methods.

Jin & El-Saawy [49] attempt to optimize a portfolio of two stocks (a high-beta stock and a low-beta stock) by using Q-Learning with a neural network approximating the state-action value function in a similar manner as the DQN (although they use a multilayer perceptron, MLP, instead of a CNN). Input features (without discretization) are the historical stock prices, number of outstanding shares in each stock, total portfolio value, and the amount of left-over cash. The agent's action is the percentage of the portfolio for which the low-beta stocks are sold and the high-beta stocks are bought, whose action space is discretized into seven percentages (+/-25%, +/-10%, +/-5%, 0%). The reward function with transaction cost considered is the portfolio return or Sharpe ratio, both penalized by the portfolio volatility. The agent is trained using both an $\varepsilon$-greedy exploration strategy and an experience replay buffer as in the DQN. The out-of-sample test result shows that the portfolio performance varies significantly depending on the performance metric, the length of stock price history, and the volatility penalty. The highest average Sharpe ratio was obtained using the portfolio return reward penalized by volatility.

Weijs [50] tackles the portfolio optimization problem also by using Q-Learning with a neural network approximating the state-action value function, with asset returns as sole input feature without discretization, and without any assumption on the return distribution. The author constructs a portfolio of a short-term T-Bill, a long-term T-Note, and a risky stock market portfolio (with transaction costs considered). Reward clipping [51] is used to rescale the rewards in the range [-1, 1]. If rewards are predominantly positive, it will result in a nearly monotonic growth of the Q-values, thereby becoming non-convergent. It is claimed that the Q-Learning method achieves in a historical out-of-sample test second-order stochastic dominance – i.e., a comparable performance with a risk-neutral investor (higher expected return) as well as a risk-averse investor (lower standard deviation). Also claimed is a lower turnover as a result of the method's conservative strategy of rarely investing fully in the assets.

## 3.3. Policy-Based Methods

### 3.3.1. Deep Deterministic Policy Gradient
In model-free and policy-based methods, the policy itself is directly parametrized with $\theta$ to represent the optimal policy $\pi^* \sim \pi_\theta$. To this end, the performance objective can be rewritten using the conditional probability density $\pi_\theta(a|s)$ as:

$$J(\pi_\theta) = \int_S \rho^\pi(s) \int_A \pi_\theta(a|s) r(s,a) \, da \, ds = \mathrm{E}_{s \sim \rho^\pi, a \sim \pi_\theta}[r(s,a)] \qquad (16)$$

where $S$ is a state space, $A$ is an action space, and $\rho^\pi$ is the discounted state distribution:

$$\rho^\pi(s') = \int_S \sum_{t=1}^\infty \gamma^t p_0(s) p(s \to s', t, \pi) ds \qquad (17)$$

where $p(s \to s', t, \pi)$ is the density at state $s'$ after transitioning from $s$ for time steps $t$. The parameters $\theta$ are then updated by performing stochastic gradient ascend in the direction of the performance gradient $\nabla_\theta J(\pi_\theta)$:

$$\begin{aligned}\nabla_\theta J(\pi_\theta) &= \int_S \rho^\pi(s) \int_A \nabla_\theta \pi_\theta(a|s) Q^\pi(s,a) da\, ds \\ &= \mathrm{E}_{s \sim \rho^\pi, a \sim \pi_\theta}[\nabla_\theta \log \pi_\theta(a|s) Q^\pi(s,a)]\end{aligned} \qquad (18)$$

(see [52] for poof), which is also called the policy gradient (i.e., the gradient of the policy's performance). In practice, $Q^\pi(s, a)$ is Monte-Carlo estimated with a sample return $R_t$ (as in the REINFORCE algorithm [53]) or replaced with an estimate of the advantage function $A(s, a) = Q(s, a) - V(s)$, computed with $R_t$ (which itself is an estimate of $Q^\pi$) and a learned estimate of $V^\pi$. The latter case can be viewed as an actor-critic algorithm [1], where the policy $\pi$ is the actor and the approximated $V^\pi$ is the critic.

Policy Gradient (PG) [52] represents the policy by a parametric probability distribution $\pi_\theta(a|s) = P(a|s; \theta)$ that selects action $a$ in state $s$ according to parameter vector $\theta$, the selection of which action can be both stochastic and deterministic. In the stochastic case (a.k.a. Stochastic Policy Gradient; SPG [54]) the policy gradient integrates over both state and action spaces as in Eq. 18, while in the deterministic case (a.k.a. Deterministic Policy Gradient; DPG [55]) it only integrates over the state space. Consequently, the SPG may require more samples than the DPG, especially if the action space has many dimensions.

The Deep Deterministic Policy Gradient (DDPG) [15] is a model-free, off-policy, and actor-critic method combining the policy-based DPG and the value-based DQN for large continuous domains, where the actor, which is a parametrized deterministic policy $\mu(s|\theta^\mu)$, learns using the Bellman equation as in Q-Learning based on feedback from the critic, which is the state-action value function $Q(s, a)$. In doing so, the algorithm trades off variance reduced by the DPG with bias introduced from Q-Learning. The actor is updated by performing gradient ascend in the direction of the policy gradient $\nabla_{\theta^\mu} J$:

$$\begin{aligned}\nabla_{\theta^\mu} J &\approx \mathrm{E}\left[\nabla_{\theta^\mu} Q(s,a|\theta^Q)|_{s=s_t, a=\mu(s_t|\theta^\mu)}\right] \\ &= \mathrm{E}\left[\nabla_a Q(s,a|\theta^Q)|_{s=s_t, a=\mu(s_t)} \nabla_{\theta^\mu} \mu(s|\theta^\mu)|_{s=s_t}\right]\end{aligned} \qquad (19)$$

(see [55] for proof). One challenge when using neural networks for RL is that most optimization algorithms assume i.i.d. samples, which assumption no longer holds when the samples are generated by sequentially exploring in an environment. Both the DQN and DDPG solve this issue by using an experience replay buffer. At each time step, the actor and critic are updated by sampling a mini-batch of uncorrelated transitions from the buffer. In addition, for efficient learning and finding hyper-parameters which generalize across environments with different scales of state values, the DDPG uses batch normalization – i.e., normalizing each feature dimension across the samples in a mini-batch to have unit mean and variance.

Jiang et al. [14] present a DDPG-like solution (without an actor-critic architecture) to the portfolio optimization problem. They use input features consisting of the highest, lowest and closing prices of portfolio component assets as continuous state, and train an agent with a Deep Neural Network (DNN) approximated policy function – what the authors call the Ensemble of Identical Independent Evaluators (EIIE; Figure 2), which evaluates each asset's potential growth in the immediate future – using the fully exploiting DPG so as to directly compute a set of portfolio weights as the agent's action. The DPG follows the gradient of the state-action value function which in their case is simply the adjusted sample return $r_t / t_f$ obtained immediately after taking an action, where $r_t$ is the portfolio's logarithmic return at time $t$, and $t_f$ is the length of the whole portfolio management period. It then determines an action within a constrained continuous space (i.e., portfolio weights with full-investment and non-negative constraints) based on the softmax voting scores from the last hidden layer in the EIIE[4]. We have noticed that their method often ends up in extreme weights (overrconcentration in very few assets) since it does not take into consideration the volatility of the portfolio (i.e., the reward function is the explicit average of the periodic logarithmic returns without any penalization by portfolio volatility), and hence it resembles the KCP portfolio (Section 2.2). Also, such extreme weights often alternate between 0 and 1 in a short period time, showing instability. Moreover, the method can be used only for buy-and-hold strategies, which is not particularly universal.

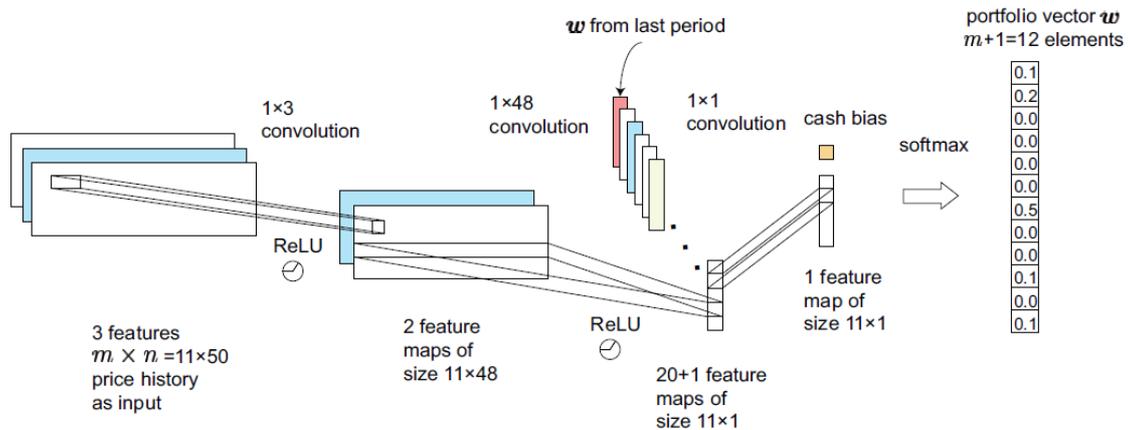

**Figure 2** (extracted from p.14 of [14]): Convolutional Neural Network (CNN) version of the Ensemble of Identical Independent Evaluators (EIIE) topology.

Guo et al. [61] uses a Convolutional Neural Network (CNN) to directly approximate the optimal policy function, where the agent's goal is to maximize the expected logarithmic return of a portfolio of 100 randomly selected stocks (i.e., log-optimal portfolio). It therefore resembles the KCP portfolio but their objective function is penalized by portfolio volatility (without transaction cost). They call this framework the Robust Log-Optimal Strategy with Reinforcement Learning (RLOSRL)[5], whose main advantage is that it does not require the complex estimation of the asset return density distributions but instead the estimated means and covariance matrix. The estimation of the two is done by pattern matching – i.e., computing the Pearson correlation coefficient between the current price returns and historical returns, and choosing the time period in which the coefficient is the highest. The input features to the network are the opening, highest, and lowest prices of each stock as well as its trading volume. The network is trained using an experience replay with a Poisson distribution sampling to emphasize the recent experience. Their out-of-sample backtests show that the RLOSRL outperforms all benchmarks in the testing period.

---

[4] An EIIE implementation in Python using TensorFlow is available at [56] under the GPL source code license.
[5] An RLOSRL implementation in Python using TensorFlow is available at [62].

### 3.3.2. Proximal Policy Optimization

Approximating the optimal policy $\pi^*$ directly with a neural network with many parameters is difficult and can often suffer from suboptimal solutions mainly due to instability and sample inefficiency. One workaround for this issue is the on-policy Proximal Policy Optimization (PPO) [17] algorithm. The PPO is an extension to the Trust Region Policy Optimization (TRPO) [57] algorithm, so the latter needs to be discussed first. The idea of TRPO is to restrict each policy gradient update, measured by the Kullback-Leibler (KL) divergence of two continuous probability distributions:

$$D_{KL}(P,Q) = \mathrm{E}\left[\log \frac{p(x)}{q(x)}\right] = \int_{-\infty}^{\infty} p(x) \log \frac{p(x)}{q(x)} dx \qquad (20)$$

to fall within a "trust region" so as to prevent updated policies from having wild deviations from previous policies. Preventing destructively large policy updates allows one to run multiple epochs of stochastic gradient ascent on the same samples, thereby reducing sample inefficiency.

Let $\eta(\theta) = J(\pi_\theta)$ be the performance of a parametrized stochastic policy $\pi_\theta$. Its lower bound function $M(\theta)$ can be defined as:

$$\eta(\boldsymbol{\theta}) \geq M(\boldsymbol{\theta}) = L(\boldsymbol{\theta}) - C \cdot KL$$
$$= \mathrm{E}\left[\frac{\pi_\theta(a|s)}{\pi_{\theta_{old}}(a|s)} \hat{A}\right] - \frac{4\varepsilon\gamma}{(1-\gamma)^2} \cdot \max_s D_{KL}\left(\pi_\theta(\,\cdot\,|s), \pi_{\theta_{old}}(\,\cdot\,|s)\right) \qquad (21)$$

(see [57] for proof), where $L(\theta)$ is the expected advantage function for the current policy calibrated by the probability ratio of the current policy to the old one. The use of advantage function instead of the expected reward $J$ reduces the variance of the estimation. Figure 3 shows the relationship among $\eta$, $M$ and $L$ at $\boldsymbol{\theta}$.

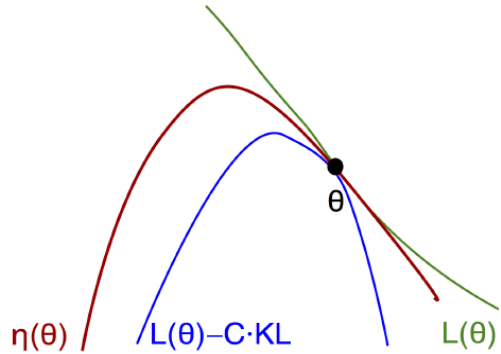

**Figure 3** (extracted from p.7 of [60]): Lower bound function $M$ (in blue).

The accuracy of $L$ approximating the advantage function locally at the current policy parameter $\boldsymbol{\theta}$ gets lower as it moves away from the old policy parameter $\boldsymbol{\theta}_{old}$, but this inaccuracy has an upper bound, which is the second term in $M$ (Eq. 21). We can therefore obtain the optimal parameter $\boldsymbol{\theta}^* = \mathrm{argmax}_\theta\, \eta(\boldsymbol{\theta})$ by applying on $M$ the Minorize-Maximization (MM) algorithm [58], in which improvements in the policy performance $\eta$ are monotonic. However, in practice it is hard to find the maximum of the KL divergence, hence it is relaxed to use the mean (i.e., $\mathrm{E}[D_{KL}]$) instead. Moreover, the penalty coefficient $C$ can hardly be used since its theoretical result would provide policy updates with step sizes that are too small. As a result, the TRPO uses $L(\theta)$ as the "surrogate" objective function which is maximized subject to a constraint on the size of the policy update:

$$\underset{\theta}{\text{maximize}} \ \mathrm{E}\left[\frac{\pi_\theta(a|s)}{\pi_{\theta_{old}}(a|s)} \hat{A}\right] \quad (22)$$

$$\text{subject to } \mathrm{E}\left[D_{KL}\left(\pi_\theta(\ \cdot\ |s), \pi_{\theta_{old}}(\ \cdot\ |s)\right)\right] \leq \delta \quad (23)$$

whose solution requires the calculation of the second-order gradient as well as its inverse, making the TRPO somewhat impractical.

The PPO, on the other hand, requires only the first-order gradient, thus is easier to implement. Let $r(\theta)$ denote the probability ratio of stochastic policies:

$$r(\theta) = \frac{\pi_\theta(a|s)}{\pi_{\theta_{old}}(a|s)} \quad (24)$$

so $r(\theta_{old}) = 1$. The PPO's unconstrained surrogate objective is then defined as:

$$L^{CLIP}(\theta) = \mathrm{E}\left[\min\left(r(\theta)\hat{A}, \text{clip}(r(\theta), 1-\varepsilon, 1+\varepsilon)\hat{A}\right)\right] \quad (25)$$

where $\varepsilon$ is a hyperparameter defining the clipping interval for $r(\theta)$.

Figure 4 illustrates the effect of this clipping. On the left graph A > 0 means the action had a positive effect on the outcome. In this case, if $r > 1 + \varepsilon$ (i.e., the action became more probable in the current policy than in the old one), the objective gets flat and thus its gradient becomes zero, preventing the stochastic gradient ascent algorithm from stepping further so as not to make the action excessively more probable, thereby making the policy stable. On the other hand, if the action became less probable when A > 0 (i.e., the policy gets worse accidentally), the gradient is still positive and we can therefore make a "correction" by ascending in the gradient's direction.

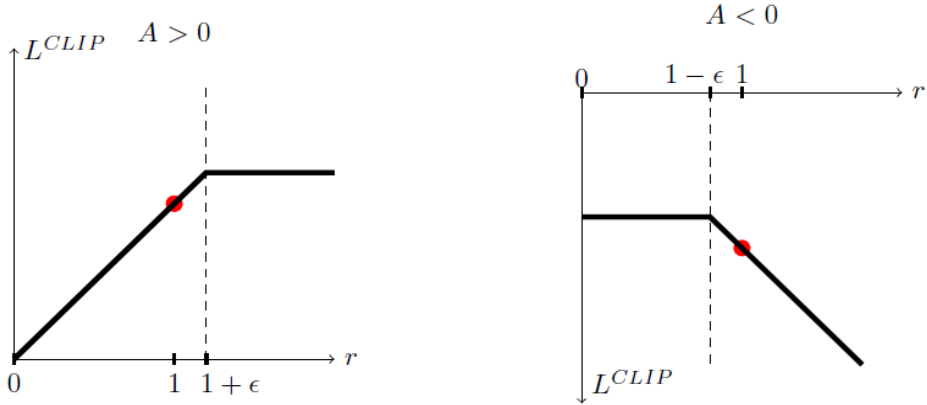

**Figure 4** (extracted from p.3 of [17]): Effects of the objective clipping in PPO.

In the same manner, on the right graph A < 0 means the action had a negative effect. If the action became less probable in the current policy, we ensure it will not get excessively less so in order to make the policy stable. But if the action became more probable when A < 0 (i.e., the policy gets worsened), the objective gets proportionately more negative, thus it has a negative gradient, with which we can revert the ascend to correct the policy.

Liang et al. [16] experiment with the DPG, DDPG, and PPO (Figure 5) for the portfolio optimization problem. They construct portfolios consisting of five randomly chosen Chinese stocks and periodically optimize the portfolio using each of the RL methods with a risk-adjusted reward function penalized by the average price volatility of the component stocks. Input features used are the highest and closing prices with random noise added to implement adversarial learning. Their DDPG also uses the EIIE topology [14] from Section 2.3.3 but employ the Deep Residual Network (ResNet) [59] instead of the CNN in order to solve the vanishing and exploding gradients problem when the depth of the networks is increased. According to the authors both the DDPG and PPO failed to learn the optimal policy even in training, hence their results are completely omitted. The DPG with adversarial learning was the only method shown successful in both learning and outperforming benchmark portfolios in backtests. The authors claim that adversarial learning increases both the daily return and Sharpe ratio of the portfolio but simultaneously increase the downside risk measured as maximum drawdown.

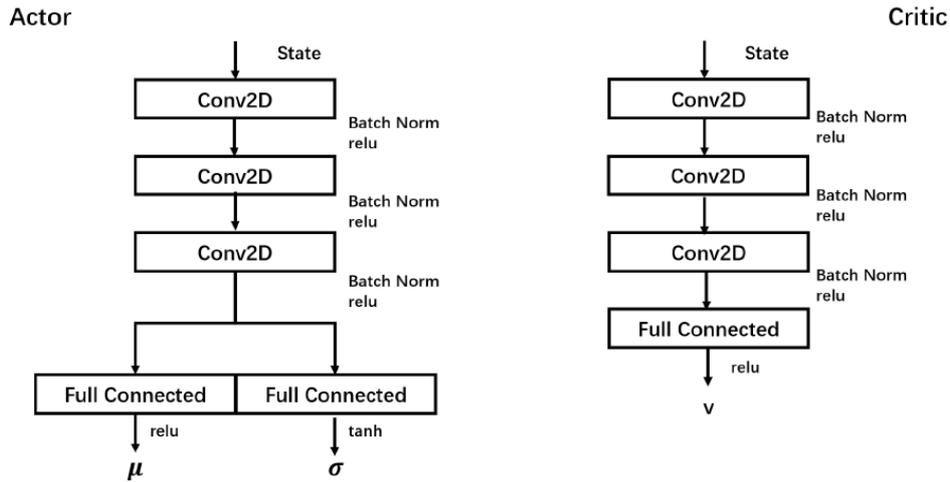

**Figure 5** (extracted from p.8 of [16]): The actor-critic PPO topology.

## 4. Conclusion and Discussion

In this survey, we have reviewed both value-based and policy-based model-free Reinforcement Learning (RL) methods applied to the portfolio optimization problem. The biggest disadvantage of the valued-based RL methods (such as Q-Learning) is Bellman's curse of dimensionality that arises from large state and action spaces, making it difficult for the agent to efficiently explore large action spaces. Du et al. [47] used descritized states and three different state-action value functions for Q-Learning. Jin & El-Saawy [49] used continuous state inputs but descritized the action space to train a neural network approximating the state-action value function. Performance of both their agents varied significantly depending on the type of the Q-value function (in the case of Du et al.) or the type of performance metric, the length of stock price history, and the volatility penalty in the reward function (in the case of Jin & El-Saawy). The variance in portfolio performance can be attributed to the fact that 1) determining the global optimal policy in an arbitrary Q-value function is often infeasible without guaranteed convexity, and 2) Q-Learning also suffers from instability for optimal policy selection when random noise exists in the loss function (Eq. 21). Such randomness can come from the stochastic dynamics or partial observability of the environment, *aleatoric* uncertainty in the reward function (including human error in labeling), or *epistemic* uncertainty in the Q-value function (e.g., poor function approximation).

The policy-based RL methods, on the other hand, can be applied directly to large continuous domains. However, approximating the optimal policy with a neural network with many parameters (including hyperparameters) is difficult and can suffer from suboptimal solutions mainly due to its instability, sample inefficiency, and sensitivity on the selection of hyperparameter values. Jiang et al. [14] constructed a DDPG-like neural network framework with continuous state inputs to directly approximate the optimal policy. Portfolio weights produced by their agent resembled those of the risky KCP portfolio that we discussed in Section 2.2, and the weights often alternated between 0 and 1 in a short period time, showing a high degree of instability. Liang et al. [16] constructed three different portfolio agents each using the DPG, DDPG, and PPO with continuous state inputs, but both the DDPG and PPO agents failed to learn optimal policy in training despite them being more advanced methods of Deep Reinforcement Learning (DRL) than the DPG.

The fact that model-free RL solves decision-making problems from the beginning – i.e., by iteratively updating Q-values or adjusting its policy parameters for each new state and action – makes the dynamics of its learning process fundamentally dependent on the samples used. In the particular case of the portfolio optimization problem, historical prices of the financial assets used for training various agents are just one realized path of the complex process of market price discovery within the specific time period in which the seemingly random realization took place. Because of this, model-free RL is particularly sample inefficient for the portfolio optimization problem (and for other decision-making problems in financial markets as well). Perhaps more importantly, it is also susceptible to overfitting as repeated training using the same historical dataset would end up capturing random patterns that exist only in that particular set of samples used. Agents trained in such settings would have null power over the future, regardless of how well they appear to have worked in the past. Also, if the neural network approximating the Q-value function or the policy function is overly complex, this also ends up in overfitting because when the network has a very high degree of freedom, all the weights and biases in the network can internally represent each individual input in the training dataset. It therefore does not generalize anything but will only replicate previously seen state-action and reward combinations. Also, the separation of historical datasets into an in-sample training set and an out-of-sample testing set will be meaningless if the out-of-sample results are repeatedly used as a feedback to update the hyperparameters of the network and to re-train the agent in the training set.

Ashby's Law of Requisite Variety [68] states that, if a system is to be stable, the number of states of its control mechanism must be greater than or equal to the number of states in the system being controlled. Experience replay, originally used in the DQN, has the advantage of reducing sequential or temporal correlations in samples, but it still cannot take into account many other possible states. Unlike model-free RL, *model-based* RL methods can simulate transitions using a learned model, resulting in increased sample efficiency and therefore increased stability. Yu et al. [63] propose a model-based RL method for the portfolio optimization problem, where they make use of a Generative Adversarial Network (GAN) [64] to generate synthetic market data to overcome sample inefficiency. Specifically, they train the Recurrent GAN (RGAN) [65] using real historical price data to produce convincingly "realistic" multidimensional time-series data (i.e., the highest, lowest, and closing prices), with which they make price predictions via the Nonlinear Dynamic Boltzmann Machine (NdyBM) [66] as well as the WaveNet [67] and train the off-policy actor-critic DDPG algorithm in an imitation learning framework (with transaction cost and slippage considered). Their model can also be used for the on-policy actor-critic PPO, which we discussed in Section 3.3.2. They show a promising out-of-sample testing result of their model, although we do not know how many times they repeated the testing to calibrate the hyperparameters and re-train the models.

All of the surveyed RL methods that use neural networks seem to be based on the assumption that the past asset return is a good predictor of the future asset return, inasmuch as they all use past historical price data as input features to their neural networks. In reality, however, (as some of them admit) asset prices behave rather independently of their past performance, and consequently the past is often not a good predictor of the future at all in the financial markets. This implies that their state spaces are an underrepresentation of the market environment, which spaces are hence insufficiently informative for the agent to learn the optimal policy. Therefore, it would be necessary to provide a more efficient and detailed state representation by using more meaningful features, e.g., fundamentals data or market sentiment data, and furthermore to investigate the predictive power of those features for future returns before attempting to apply them on the equally complex problem of portfolio optimization, the former of which tasks would be difficult given the very low signal-to-noise ratio of many financial datasets and the non-ergodic nature of the financial markets. For this reason, we believe that domain knowledge still plays an important role in feature engineering and selection, despite DRL's low requirements for feature engineering. Moreover, those features should be able to sufficiently predict the return of different kinds of investment strategies, not just the buy-and-hold strategy, so as to make the portfolio more universal.

As Weijs [50] points out, the interpretability of machine learning models is another important issue in applying RL to the portfolio optimization problem, since institutional investors do not want to risk a large amount of capital in a model that cannot be explained by financial or economic theories, nor in a model for which the human portfolio manager cannot be responsible. Deep Neural Networks (DNNs) are notorious for being a "black box" as their hidden layers exhibit many-to-many complex relationships. In DRL, the optimal policy must be inferred by the agent's trial-and-error interaction with the environment, in which the agent is fundamentally driven by the DNN black box, and the only learning signal the agent receives is the scalar reward. Reward functions are difficult to design and hard to get to work in many problems. If they are not specified properly in advance, the agent could end up being caught in local minima, causing unintended and unpredicted behaviors, which could potentially cause a significant amount of financial loss in actual portfolio management.

Last but not least, the (infamous) credit assignment problem in RL – i.e., the situation wherein the consequences of the agent's actions only materialize after many transitions of the environment – is another issue in the portfolio optimization problem. Although the actions are always aimed at maximizing the (risk-adjusted) return of the portfolio at a certain rebalancing frequency (in other words, the timeframe of credit assignment is clearly defined), the structure of credit assignment can change over time due to the non-ergodicity of the financial markets, which brings *out-of-distribution* uncertainty, potentially causing the agent to merely learn (in hindsight) a random policy.

# 5. References

[1] Sutton, R.G. and Barto, A.G. Reinforcement Learning: An Introduction. *MIT Press*, 1998.
[2] Mnih, V., Kavukcuoglu, K., et al. Human-Level Control through Deep Reinforcement Learning. *Nature*, 518(7540):529–533, 2015.
[3] Hessel, M., Modayil, J, et al. Rainbow: Combining Improvements in Deep Reinforcement Learning. *ArXiv*:1710.02298, 2017.
[4] Silver, D., Huang, A., et al. Mastering the Game of Go with Deep Neural Networks and Tree Search. *Nature*, 529(7587):484–489, 2016.
[5] Silver, D., Hubert, T., et al. A general reinforcement learning algorithm that masters chess, shogi, and Go through self-play. *Science*, 362, 1140–1144, 2018.
[6] Gervais, A., Karame, G.O., et al. On the Security and Performance of Proof of Work Blockchains. *Blockchain Protocol Analysis and Security Engineering*, January 2017.

[7] Eyal, I., and Sirer. E.G. Majority is not enough: Bitcoin mining is vulnerable. *Financial Cryptography and Data Security*, pages 436–454. Springer, 2014.

[8] Nakamoto, S. Bitcoin: A Peer-to-Peer Electronic Cash System. 2008.

[9] Halperin, I. QLBS: Q-Learner in the Black-Scholes(-Merton) Worlds. *ArXiv*:1712.04609, 2017.

[10] Black, F. and Scholes, M. The Pricing of Options and Corporate Liabilities. *Journal of Political Economy*, Vol. 81(3), 637-654, 1973.

[11] Merton, R. Theory of Rational Option Pricing. *Bell Journal of Economics and Management Science*, Vol.4(1), 141-183, 1974.

[12] Watkins, C.J. and Dayan, P. Q-Learning, *Machine Learning*, 8(3-4), 179-192, 1992.

[13] Buehler, H., Gonon, L., et al. Deep Hedging. *ArXiv*:1802.03042, 2018.

[14] Jiang, Z., Xu, J.D., et al. A Deep Reinforcement Learning Framework for the Financial Portfolio Management Problem. *ArXiv*:1706.10059, 2017.

[15] Lillicrap, T.P., Hunt, J.J., et al. Continuous Control with Deep Reinforcement Learning. *ArXiv*:1509.02971, 2016.

[16] Liang, Z., Chen, H., et al. Adversarial Deep Reinforcement Learning in Portfolio Management. *ArXiv*:1808.09940, 2018.

[17] Schulman, J., Wolski, F., et al. Proximal Policy Optimization Algorithms. *ArXiv*:1707.06347, 2017.

[18] Baz, J. and Guo, H. An Asset Allocation Primer: Connecting Markowitz, Kelly and Risk Parity. *PIMCO*, 2017.

[19] Markowitz, H.M. Portfolio Selection. *Journal of Finance*, 7(1), 77–91, 1952.

[20] Markowitz, H.M. The Optimization of a Quadratic Function Subject to Linear Constraints. *Naval Research Logistics Quarterly*, III, 111–133, 1956.

[21] Markowitz, H.M. Portfolio Selection: Efficient Diversification of Investments. *John Wiley and Sons*, 1959.

[22] Meucci, A. Risk and Asset Allocation. *Springer*, 2005.

[23] Michaud, R. Efficient asset allocation: A practical guide to stock portfolio optimization and asset allocation. *Harvard Business School Press*, 1998.

[24] Bailey, D.H. and López de Prado, M. Balanced Baskets: A new approach to Trading and Hedging Risks. *Journal of Investment Strategies*, Vol. 1, No. 4, pp. 21-62, 2012.

[25] López de Prado, M. Building Diversified Portfolios that Outperform Out-of-Sample. *Journal of Portfolio Management*, 2016.

[26] Bailey, D.H. and López de Prado, M. An Open-Source Implementation of the Critical-Line Algorithm for Portfolio Optimization. *Algorithms*, 6, 169–196, 2013.

[27] Kelly, J.L. A new interpretation of information rate. *Bell System Technical Journal*, pages 917–926, 1956.

[28] Thorp, E. The Kelly Criterion in Blackjack, Sports Betting and the Stock Market. *Handbook of Asset and Liability Management, North Holland*, 1, 385-428, 2006.

[29] Ethier, S.N. The Kelly system maximizes median fortune. *Journal of Applied Probability*, 41(5):1230 – 1236, 2004.

[30] Breiman, L. Optimal gambling systems for favorable games. *Proc. 4th Berkeley Symp. Math. Statist. Prob.* 1, 65-78, 1961.

[31] Bell, R.M. and Cover, T.M. Competitive Optimality of Logarithmic Investment. *Mathematics of Operations Research*, 5(2):161-166, May 1980.

[32] Bochman, A. Half of what you've read about the Kelly Criterion is wrong. 2018. https://www.linkedin.com/pulse/half-what-youve-read-kelly-criterion-wrong-alon-bochman-cfa/

[33] Bellman, R. and Kalaba, R. On the role of dynamic programming in statistical

[33] communication theory. *Information Theory, IRE Transactions*, 3(3):197–203, 1957.

[34] Mossin, J. Optimal multiperiod portfolio policies. *The Journal of Business*, 41(2):215–229, 1968.

[35] Nekrasov, V. Kelly criterion for multivariate portfolios: a model-free approach. *SSRN*, 2014.

[36] Ziemba, W.T. A response to Professor Paul A. Samuelson's objections to Kelly capital growth investing, *Journal of Portfolio Management*, 42(1): 153–67, 2015.

[37] Laureti, P., Medo, M., et al. Analysis of Kelly-optimal portfolios. *ArXiv*:0712.2771, 2009.

[38] Markowitz, H.M. Investment in the Long Run: New Evidence for an Old Rule. *Journal of Finance*, 31, 1273–1286, 1976.

[39] Cornuejols, G. and Tutuncu, R. Optimization Methods in Finance. Page 141, *Carnegie Mellon University*, January 2006.

[40] Wikipedia. Taylor series.
https://en.wikipedia.org/wiki/Taylor_series#Natural_logarithm

[41] Dama, M. Max Dama on Automated Trading. Page 36, 2011.
http://isomorphisms.sdf.org/maxdama.pdf

[42] Lo, A.W. Adaptive Markets: Financial Evolution at the Speed of Thought. *Princeton University Press*, page 240, 2017.

[43] Merton, R.C. On Estimating the Expected Return on the Market An Exploratory Investigation. *Journal of Financial Economics*, Vol. 8, pp. 323-361, 1980.

[44] Lee W. Risk-Based Asset Allocation: A New Answer to an Old Question? *Journal of Portfolio Management*, 37, 11-28, 2011.

[45] Melo, F.S. Convergence of q-learning: A simple proof. *Institute Of. Systems and Robotics, Tech. Rep*, pp. 1–4, 2001.

[46] Jaakkola, T., Jordan, M.I., et al. Convergence of stochastic iterative dynamic programming algorithms. *Neural Computation*, 1994.

[47] Du, X., Jinjian Z., et al. Algorithm trading using q-learning and recurrent reinforcement learning. *Positions*, 1, 2009.

[48] Moody, J., Wu, L.,et al. Performance functions and reinforcement learning for trading systems and portfolios. *Science*, 17 (February 1997), 441–470, 1998.

[49] Jin, O. and El-Saawy, H. Portfolio Management usig Reinforcement Learning. *Stanford University*, 2016.

[50] Weijs, L. Reinforcement learning in Portfolio Management and its interpretation. *Erasmus Universiteit Rotterdam*, 2018.

[51] Hasselt, H., Guez, A., et al. Learning values across many orders of magnitude. *ArXiv*:1602.07714, 2016.

[52] Sutton, R.S., McAllester, D.A., et al. Policy gradient methods for reinforcement learning with function approximation. *Neural Information Processing Systems*, 12, pages 1057–1063, 1999.

[53] Williams, R.J. Simple statistical gradient following algorithms for connectionist reinforcement learning. *Machine Learning*, 8:229–256, 1992.

[54] Degris, T., Pilarski, P. M., et al. Model-free reinforcement learning with continuous action in practice. *American Control Conference*, 2012.

[55] Silver, D., Lever, G., et al. Deterministic Policy Gradient Algorithms. *In Proceedings of the 31st International Conference on Machine Learning (ICML-14)*, pages 387–395, 2014.

[56] Jiang, Z. PGPortfolio. Github, 2017.
https://github.com/ZhengyaoJiang/PGPortfolio

[57] Schulman, J., Levine, S., et al. Trust Region Policy Optimization. *ArXiv*:1502.05477, 2017.

[58] Wikipedia. MM algorithm. https://en.wikipedia.org/wiki/MM_algorithm.

[59] He, K., Zhang, X., et al. Deep residual learning for image recognition. *In Proceedings of the IEEE conference on computer vision and pattern recognition*, 770-778, 2016.

[60] Schulman, J. Advanced Policy Gradient Methods: Natural Gradient, TRPO, and More. *OpenAI*, 2017.

[61] Guo, Y., Fu, X., et al. Robust Log-Optimal Strategy with Reinforcement Learning. *ArXiv*:1805.00205, 2018.

[62] Fu, X. Robust-Log-Optimal-Strategy-with-Reinforcement-Learning. *Github*, 2018. https://github.com/fxy96/Robust-Log-Optimal-Strategy-with-Reinforcement-Learning

[63] Yu, P., Lee, J.S., et al. Model-based Deep Reinforcement Learning for Dynamic Portfolio Optimization. *ArXiv*:1901.08740, 2019.

[64] Goodfellow, I., Pouget-Abadie, J., et al. Generative adversarial nets. *Advances in neural information processing systems*, pp. 2672–2680, 2014.

[65] Esteban, C., Hyland, S. L., et al. Real-valued (medical) time series generation with recurrent conditional GANs. *arXiv*:1706.02633, 2017.

[66] Dasgupta, S. and Osogami, T. Nonlinear dynamic boltzmann machines for time-series prediction. *AAAI*, pp. 1833–1839, 2017.

[67] van den Oord, A., Li, Y., et al. Parallel WaveNet: Fast high-fidelity speech synthesis. In Dy, J. and Krause, A. (eds.), *In Proceedings of the 35th International Conference on Machine Learning*, volume 80 of Proceedings of Machine Learning Research, pp. 3918–3926, 2018.

[68] Ashby, W.R. Requisite variety and its implications for the control of complex systems. *Facets of systems science*, pp. 405-417, Springer, 1991.